# Surface-Plasmon Excitation of Second-Harmonic light: Emission and Absorption

MARIA A. VINCENTI[*,1], DOMENICO DE CEGLIA[1], COSTANTINO DE ANGELIS[2], MICHAEL SCALORA[3]

[1]*National Research Council – AMRDEC, Charles M. Bowden Research Laboratory, Redstone Arsenal – AL, 35898 (USA)*
[2]*Department of Information Engineering, University of Brescia, Via Branze 38, 25123 Brescia, Italy*
[3]*Charles M. Bowden Research Laboratory, AMRDEC, US Army RDECOM, Redstone Arsenal - AL, 35898 (USA)*
**Corresponding author: vincentiantonella@gmail.com*

**Abstract**

**We aim to clarify the role that absorption plays in nonlinear optical processes in a variety of metallic nanostructures, and show how it relates to emission and conversion efficiency. We define a figure of merit that establishes the structure's ability to either favor or impede second-harmonic generation. Our findings suggest that, despite the best efforts embarked upon to enhance local fields and light-coupling via plasmon excitation, nearly always the absorbed harmonic energy far surpasses the harmonic energy emitted in the far-field. Qualitative and quantitative understanding of absorption processes is crucial in the evaluation of practical designs of plasmonic nanostructures for the purpose of frequency mixing.**

## 1. Introduction

Metals do not possess a dipole-allowed quadratic nonlinear source. Yet, they have been investigated for nonlinear frequency conversion since the early days of nonlinear optics [1-9]. Second-order nonlinear contributions are due to a combination of symmetry breaking at the surface, magnetic dipoles due to the Lorentz force, inner-core electrons, convective nonlinear sources, and electron gas pressure [10-12]. Metals are also known to possess a large and relatively fast third-order nonlinearity [13, 14], and have been extensively explored for their exceptional ability to promote field enhancement in sub-wavelength regions as a result of surface-plasmon excitation [15, 16]. In this context metals can enhance transmission through sub-wavelength features in metal gratings [17-23]; promote efficient sensing [24-28]; enhance Raman scattering [29-32] and signal processing at the nanoscale [33]. High field localization and small modal volumes lend surface plasmons well to the investigation of frequency-mixing in metamaterials [34-36]; metallic gratings [12, 37-42]; layered metal-dielectric photonic band gap structures [43, 44]; and nanoantennas [45-47], to name just a few configurations. These studies have indeed corroborated the idea that the excitation of surface plasmons is systematically associated with efficient harmonic generation. Nevertheless, these investigations generally neglect absorption processes, thus providing only a limited, partial view of the problem.

Our goal in this paper is to quantify the absorption of second harmonic (SH) light relative to the generated signal. We will examine several metallic configurations that include a silver mirror, an isolated thin silver film and in the Kretschmann configuration. We will also analyze more complex but representative metallic nanostructures, such as single and coupled nanorods, and metal gratings with either grooves or slits. Notwithstanding the improvements in SH conversion efficiency, the comparison suggests that SH absorption increases dramatically when surface plasmons are more efficiently coupled to the fundamental-frequency (FF) field. This is generally understandable, since both SH conversion efficiency and absorption are proportional to the local field intensity. At the same time, almost paradoxically, SH emission may be favored over SH absorption if certain resonant conditions at the pump and SH frequencies occur. For this purpose, it may be beneficial to define a figure of merit to help one evaluate in practical terms the ability of these structures to either benefit or inhibit SH emission. The results show that, for the most part, absorbed harmonic energy can be much larger compared to the energy emitted in the far-field. These details are important when designing metallic nanostructures for frequency-mixing purposes.

## 2. Second order nonlinear processes from metallic nanostructures

### A. Metal-Dielectric interfaces

We selected a number of well-known platforms and compared their ability to radiate and absorb SH light. First we tackle the linear electromagnetic problems at the FF using a frequency-domain, finite-element solver (COMSOL Multiphysics). Then we adopt the undepleted pump approximation and use the calculated FF fields to define the SH current sources. The SH electromagnetic problem is solved by using the same finite-element solver, as outlined in Refs. [48, 49]. Specifically, SH current density sources may be calculated as the superposition of two terms: a volume term, $\mathbf{J}_{\text{vol}}$, and a surface term, $\mathbf{J}_{\text{surf}}$. These currents can then be linked to the FF electric field and to the free electron hydrodynamic parameters as follows [48, 50, 51]:

$$\hat{\mathbf{n}}\cdot\mathbf{J}_{\text{surf}} = i\frac{n_0 e^3}{2m_*^2}\frac{3+\varepsilon_{\text{FF}}}{(\omega+i\gamma_0)^2(2\omega+i\gamma_0)}E_{\text{FF},\perp}^2, \tag{1}$$

$$\hat{\mathbf{t}}\cdot\mathbf{J}_{\text{surf}} = i\frac{2n_0e^3}{m_*^2}\frac{1}{\left(\omega+i\gamma_0\right)^2\left(2\omega+i\gamma_0\right)}E_{\text{FF},\perp}E_{\text{FF},//}\,, \tag{2}$$

$$\mathbf{J}_{\text{vol}} = \frac{n_0e^3}{m_*^2}\frac{1}{\omega\left(\omega+i\gamma_0\right)\left(2\omega+i\gamma_0\right)}\left[\frac{\gamma_0}{\omega+i\gamma_0}\left(\mathbf{E}_{\text{FF}}\cdot\nabla\right)\mathbf{E}_{\text{FF}}-\frac{i}{2}\nabla\left(\mathbf{E}_{\text{FF}}\cdot\mathbf{E}_{\text{FF}}\right)\right], \tag{3}$$

where $n_0$= 4.96×10$^{28}$ m$^{-3}$ is the free electrons density in silver, the effective electron mass is $m_*$ = $m_e$ = 9.11×10$^{-31}$ Kg, $e$ is the elementary charge, $\gamma_0$ = 7.28×10$^{13}$ s$^{-1}$ is the electron gas collision frequency in silver, $\varepsilon_{\text{FF}}$ is the relative permittivity of bulk-silver at the FF, $\omega$ is the angular frequency of the FF field, $\mathbf{E}_{\text{FF}}$ is the FF electric field phasor, and $\hat{\mathbf{n}}$ and $\hat{\mathbf{t}}$ are unit vectors pointing in directions outward normal and tangential to the metallic surface, respectively. Silver permittivity data [52] are interpolated with one Drude and five Lorentz oscillators [53]. Moreover, $E_{\text{FF},\perp}$ and $E_{\text{FF},//}$ are the normal and tangential components of the FF electric field in the local boundary coordinate system defined by $\hat{\mathbf{n}}$ and $\hat{\mathbf{t}}$, respectively, and are evaluated inside the metal region. We adopt the local approximation, a safe assumption in the structures under investigation since the main effects associated with the free electrons nonlocality are perturbations of plasmonic resonances and limitations of field enhancement in sub-nanometer gaps [54-56]. In what follows we assume a pump signal tuned at $\lambda_{\text{FF}}$ = 1064 nm with irradiance $I_{\text{FF}}$ = 1 GW/cm$^2$. All structures are excited with TM-polarized light, i.e., the electric field vector lies on the *y-z* plane (see Fig. 1 and subsequent), incident at an angle $\vartheta_i$. SH far-field emission efficiency and absorption are calculated as $\eta_{\text{SH}} = \text{P}_{\text{SH}}/\text{P}_{\text{FF}}$ and $\alpha_{\text{SH}} = \text{A}_{\text{SH}}/\text{P}_{\text{FF}}$, respectively, where $\text{P}_{\text{FF}}$ is the input pump power, $\text{P}_{\text{SH}}$ is either the radiated (reflected and transmitted) or scattered power, and $\text{A}_{\text{SH}}$ is the absorbed power at the SH defined as:

$$\text{A}_{\text{SH}} = \frac{\frac{1}{2}\varepsilon_0\omega_{\text{SH}}\ \text{Im}\left[\varepsilon_{\text{SH}}\right]\int_{vol}\left\|\mathbf{E}_{\text{SH}}\right\|^2 dr}{\text{P}_{\text{FF}}}\,, \tag{4}$$

where $\omega_{\text{SH}}$ is the SH frequency, $\varepsilon_{\text{SH}}$ and $\mathbf{E}_{\text{SH}}$ are the permittivity value and the electric field at the SH frequency, respectively. The norm of $\mathbf{E}_{\text{SH}}$ is integrated over the volume of the metallic structure.

We begin by calculating the linear response of the simplest possible metal-dielectric interface, i.e., a planar silver mirror surrounded by air [Fig. 1(a)]. As shown in Fig. 1(b), reflection from the mirror shows a shallow minimum for $\vartheta_i$ ~ 77°, the pseudo-Brewster angle of the interface [57, 58]. The reflection minimum corresponds to a maximum of the generated (blue, solid line) and absorbed (red, dashed line) SH signal [Fig. 1(c)]. The absorbed SH signal is only slightly smaller than the radiated SH. A similar scenario occurs for a 30nm-thick silver layer illuminated at oblique incidence [Fig. 1(d)]. Indeed the thin film shows a deeper reflection minimum that corresponds to a transmission maximum of ~28% [Fig. 1(e)].

For the nonlinear problem we once again find that the absorbed SH signal is somewhat smaller but comparable to the total emitted (transmitted plus reflected) signal [Fig. 1(f)], and note that in the absence of any surface-plasmon excitation the absorbed SH signal is already comparable to the radiated SH. We now consider a thin silver film in the Kretschmann configuration. When the incident light impinges at the phase-matching angle, a surface plasmon is generated at the metal-air interface [Fig. 1(g)]. Typically this kind of coupling causes a sharp dip in the reflected signal [15] and a maximum peak in absorption (not shown). The coupling may be triggered by simply varying the angle of incidence of the pump [Fig. 1(h)]. Upon evaluation of the second-order nonlinear response we find that now SH absorption exceeds SH emission by a factor of two: the excitation of a propagating, surface-plasmon polariton thus favors SH absorption.

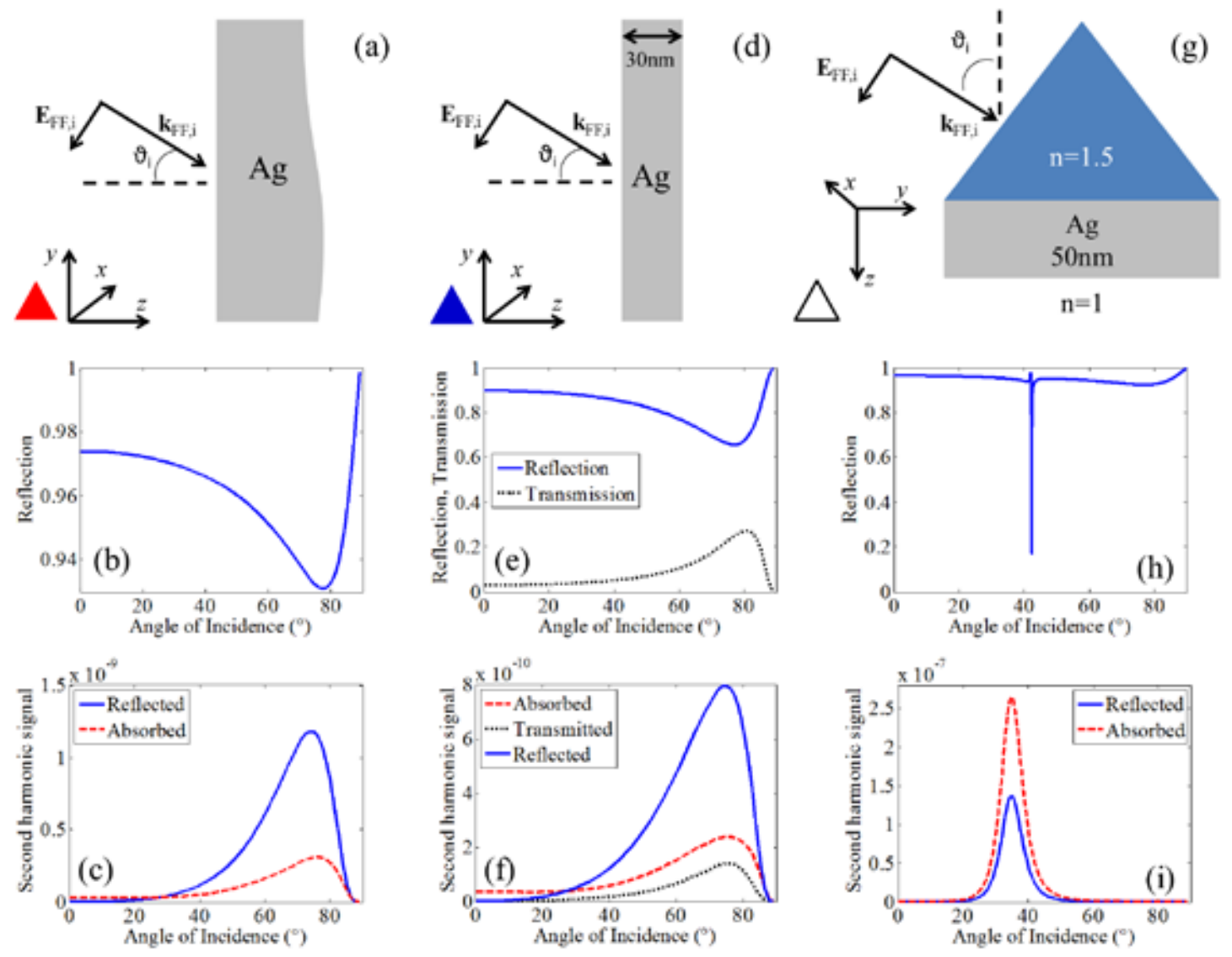

Figure 1. (a) A TM-polarized pump with electric field $\mathbf{E}_{FF,i}$ and wavevector $\mathbf{k}_{FF,i}$ impinges on a silver mirror with angle $\vartheta_i$ ; (b) Reflection vs. angle of incidence $\vartheta_i$ for the structure in (a); and (c) Reflected (blue, solid line) and absorbed (red, dashed line) SH conversion efficiency vs. angle of incidence $\vartheta_i$ for the structure in (a); (d) Same as in (a) for a thin silver layer; (e) Reflection (blue, solid line) and transmission (black, dotted line) vs. angle of incidence $\vartheta_i$ for the structure in (d); and (f) Reflected (blue, solid line), transmitted (black, dotted line), and absorbed (red, dashed line) SH conversion efficiency vs. angle of incidence $\vartheta_i$ for the structure in (d); (g) Same as in (a) for a metal layer in the Kretschmann configuration. Pump impinges on a prism with refractive index n=1.5 with angle $\vartheta_i$; (h) Reflection vs. angle of incidence $\vartheta_i$ for the structure in (g); and (i) Reflected (blue, solid line), and absorbed (red, dashed line) SH conversion efficiency vs. angle of incidence $\vartheta_i$ for the structure in (g).

**B. Single, coupled and arrayed nanorods**

The preliminary analysis in the previous section shows that absorbed and emitted SH signals are actually comparable even when excitation of surface plasmons is absent. However, nonlinear absorption surpasses far-field radiation when surface plasmons are excited in the simplest case of a thin metal film in the Kretschmann configuration. We now examine isolated metallic nanorods (or nanoantennas) that support localized surface-plasmon resonances, and then move to slightly more complex systems, such as coupled nanorods and nanorod arrays. We assume the incident field is TM-polarized and vary the incident pump wavelength. Pump irradiance is set at $I_{FF}$ = 1 GW/cm$^2$ and the structures have been designed to display a resonance around $\lambda$ = 1064 nm, so that all the structures can be fairly compared in terms of conversion efficiencies. The angle of incidence $\vartheta_i$ is set to 30°.

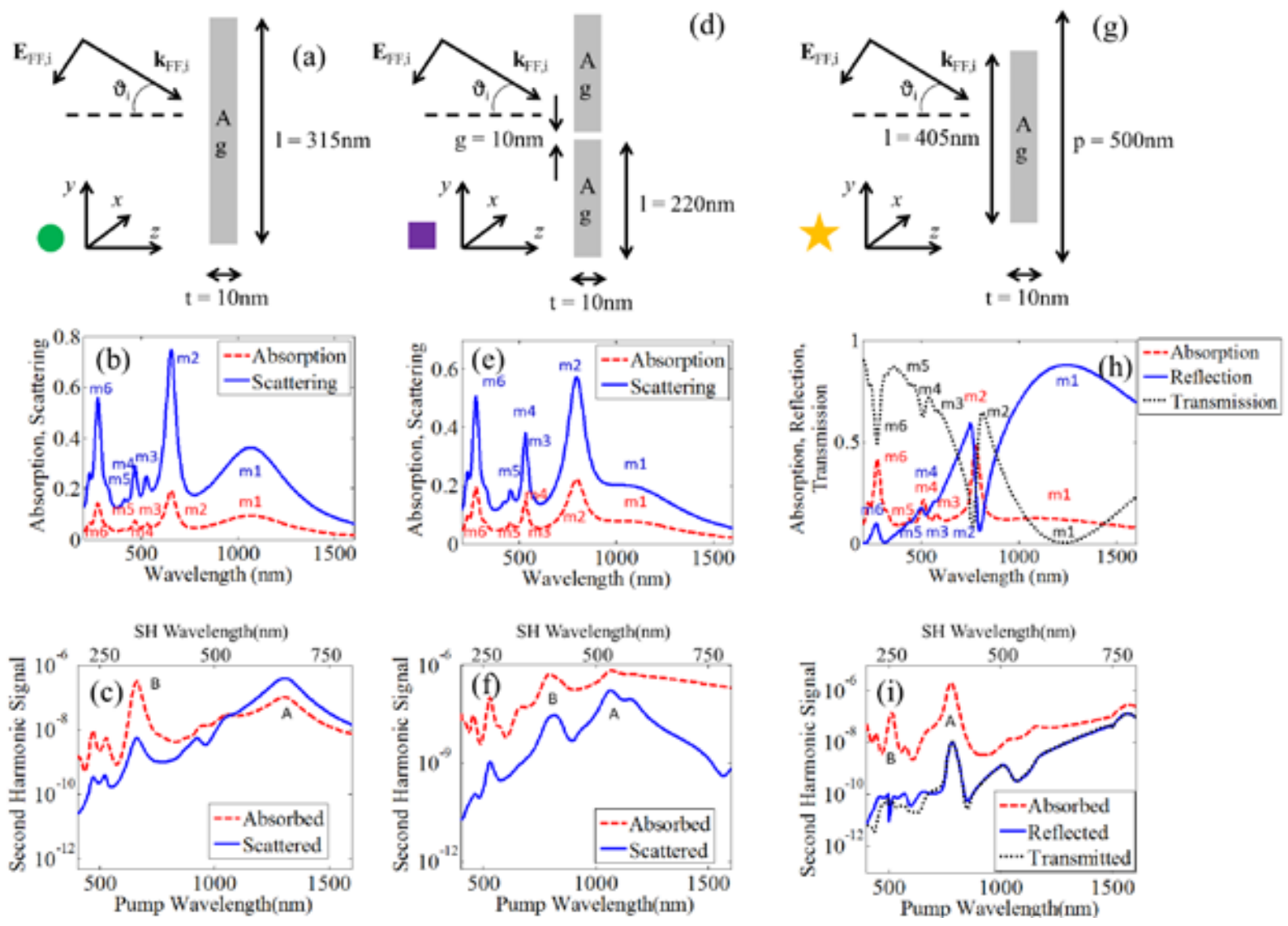

Figure 2. (a) A TM-polarized pump with electric field $\mathbf{E}_{FF,i}$ and wavevector $\mathbf{k}_{FF,i}$ impinges at $\vartheta_i = 30$ ° on a silver nanrod t=10nm thick and l=315nm long; (b) Scattering (blue, solid line) and absorption (red, dashed line) vs. incident wavelength for the structure in (a); and (c) Scattered (blue, solid line) and absorbed (red, dashed line) SH conversion efficiency vs. pump/SH wavelengths for the structure in (a); (d) Same as in (a) for two coupled nanorods separated by a gap g = 10 nm. Antennas thickness is t=10nm and length is l=220nm; (e) Scattering (blue, solid line), and absorption (red, dashed line) vs. incident wavelength for the structure in (d); and (f) Scattered (blue, solid line) and absorbed (red, dashed line) SH conversion efficiency vs. pump wavelength for the structure in (d); (g) Same as in (a) for an array of silver nanorods t=10nm thick and l=405nm long, with periodicity p=500nm; (h) Reflection (blue, solid line), transmission (black, dotted line) and absorption (red, dashed line) vs. incident wavelength for the structure in (g); and (i) Reflected (blue, solid line), transmitted (black, dotted line) and absorbed (red, dashed line) SH conversion efficiency vs. pump/SH wavelengths for the structure in (g).

Although the choice of the incident angle is not unique, oblique incidence is necessary to break the symmetry and excite both dipolar and quadrupolar modes in this kind of structures [47, 59].

We first consider a single silver nanorod of length l = 315 nm and thickness t = 10 nm, illuminated by a pump field tuned between 200nm and 1600nm [Fig. 2(a)]. The structure presents several resonant features across the entire wavelength range under investigation [Fig. 2(b)]: each resonance is associated with the excitation of a localized surface plasmon. More specifically, an isolated nanorod illuminated at oblique incidence excites both dipole-like odd modes [labeled as m1, m3, etc. in Fig. 2(b)] and quadrupole-like even modes [labeled as m2, m4, etc. in Fig.2(b)]. The oblique incidence illumination breaks the symmetry of the structure and allows coupling to both odd and even modes. In contrast, given the nature of the SH sources, more efficient SH radiation is usually obtained *via* coupling of SH light to quadrupolar modes [47]. Indeed the nonlinear analysis, carried out by sweeping the FF from 400nm to 1600nm, shows that the scattered SH signal surpasses the SH absorption when the dominant dipolar mode (m1) excited at the FF couples to the dominant quadrupolar mode (m2) excited at the SH. This process is described by the interaction m1+m1→m2 [point A Fig. 2(c)], following the schematic nomenclature introduced in Ref. [60]. In contrast, the absorbed SH signal remains dominant over SH emission for all other interactions by approximately two orders of magnitude, including coupling between quadrupolar modes at the pump and SH frequencies [see for example point B Fig. 2(c) in which the interaction is of type m2+m2→m6].

Next, we consider two coupled nanorods l=220nm long, t=10nm thick, separated by a gap g = 10nm [Fig. 2(d)]. We stress that this gap sizes allows one to assume that nonlocal and quantum effects can be neglected, so that the local approximation may be safely adopted for our calculations [54-56]. Angle of incidence of the pump is $\vartheta_i = 30°$. Multiple resonances associated with dipole-like odd modes and quadrupole-like even modes are once again visible in both scattered and absorbed linear spectra [Fig. 2(e)]. However, the nonlinear behavior of the coupled nanorods is substantially different from the nanorod: the absorbed SH signal now exceeds the scattered SH signal by one to two orders of magnitude throughout the entire spectrum [Fig. 2(f)], regardless of the kind of resonance excited at the pump frequency. A similar ratio of scattered vs. absorbed SH signal is in fact observed for both point A (m1+m1→m5) and B (m2+m2→m5) in the nonlinear spectra [Fig. 2(f)]. An even more dramatic scenario is presented when an array of nanorods is investigated [Fig. 2(g)]. Here we assume nanorods having length l = 405 nm, thickness t = 10 nm and set a periodicity p = 500 nm. These geometrical parameters have been chosen in order to obtain the fundamental resonance at $\lambda = 1064$ nm. Transmission, reflection and absorption spectra are shown in Fig. 2(h), while the nonlinear response of the structure is reported in Fig. 2(i). Once again multiple resonances (dipolar – odd modes, and quadrupolar – even modes) characterize the nanostructure. The absorbed SH signal in the visible range exceeds SH emission by approximately three orders of magnitude, and by one to two orders of magnitude above 1000 nm. For example for point A [Fig. 2(i), m2+m2→m5] absorbed SH is two orders of magnitude higher than the radiated SH, whereas for point B [Fig. 2(i), m4+m4→m6] the absorbed SH is three orders of magnitude higher than the radiated SH.

**C. Metallic gratings**

Differently from what we have seen in the previous section, where localized surface plasmons are excited, metallic gratings may support surface plasmon *via* either Fabry-Perót or Fano resonances [22, 61]. We calculate the linear and nonlinear responses of three configurations: (i) a metallic grating with shallow grooves; (ii) a metallic grating with sub-wavelength slits and periodicity comparable to the incident pump wavelength; and (iii) a metallic grating with sub-wavelength slits and periodicity. The results here once again confirm that the excitation of surface-plasmon resonances may not be the best way to enhance the magnitude of the emitted SH signal over the corresponding absorption. All structures were designed to display a resonant feature near $\lambda = 1064$ nm, and are illuminated with a TM-polarized light of variable frequency, pump irradiance $I_{FF} = 1$ GW/cm$^2$, and angle of incidence $\vartheta_i = 30°$.

The silver grating with shallow grooves (w = 45 nm deep and a = 300 nm wide) [Fig. 3(a)] and periodicity p = 700 nm supports Fano resonances associated with its periodicity [f1 and f2 in Fig. 3(b)]. The nonlinear behavior is strongly influenced by the spectral modulation due to the presence of a Fano resonance at the pump frequency [point A Fig. 3(c)] or at the SH frequency [point B Fig. 3(c)]. More specifically, pumping at the Fano resonance (point A) or exciting a Fano resonance at the SH (point B) leads to similar levels of radiated and absorbed SH energy. In contrast, for all other resonances and the remaining of the spectrum under investigation the absorbed SH energy is between one and three orders of magnitude above the radiated SH energy.

Similar results are found for a silver grating having the same periodicity p = 700 nm and sub-wavelength slits carved on a t = 200 nm-thick film [structure sketched in Fig. 3(d)]. One may observe several resonant features in the linear spectra [Fig. 3(e)], which are associated with the proximity of a Fano-like resonance to a Fabry-Perót resonance. This structure forms a plasmonic band gap due to the interference between the Fabry-Perót resonance localized inside slits and the collective resonance (Fano-like) of the grating related to the excitation of the surface plasmons at the input and output metal-air interfaces [22]. In particular, in Fig. 3(e) we highlight the resonances located on the red and blue sides of the first- (r1 and r1*, respectively) and second-order (r2 and r2*, respectively) plasmonic band gaps and the Fano-like resonances that interact with them (f1 and f2, respectively). For this structure we highlight two points in the nonlinear spectra: one point located on the red-side of the plasmonic band gap [point A Fig. 3(f)] displays a predicted SH absorption at least two orders of magnitude larger than the emitted SH signal. This behavior can be ascribed to the presence of a Fabry-Perót resonance for the slits. On the other hand, and similarly to the shallow groove system, the presence of a Fano resonance appears to favor SH light emission while mitigating SH absorption [point B in Fig. 3(f)].

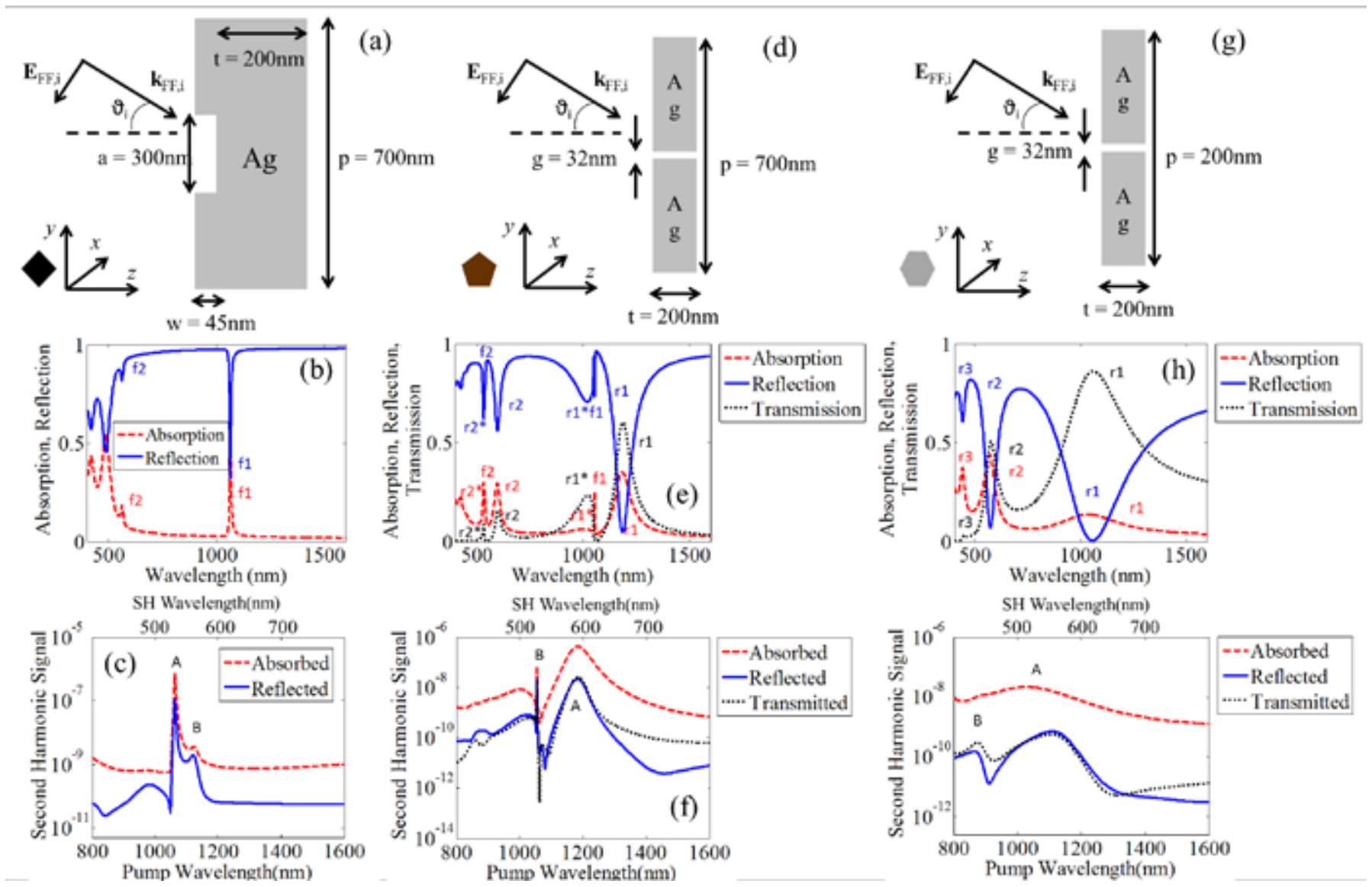

Figure 3. (a) A TM-polarized pump with electric field $\mathbf{E}_{FF,i}$ and wavevector $\mathbf{k}_{FF,i}$ impinges on a metal grating with shallow grooves at oblique incidence $\vartheta_i = 30°$; (b) Reflection (blue, solid line) and absorption (red, dashed line) vs. incident wavelength for the structure in (a); and (c) Reflected (blue, solid line) and absorbed (red, dashed line) SH conversion efficiency vs. pump/SH wavelengths for the structure in (a); (d) Same as in (a) for a metal grating with periodicity comparable to the incident wavelength and sub-wavelength slits; (e) Reflection (blue, solid line), transmission (black, dotted line) and absorption (red, dashed line) vs. incident wavelength for the structure in (d); and (f) Reflected (blue, solid line), transmitted (black, dotted line) and absorbed (red, dashed line) SH conversion efficiency vs. pump/SH wavelengths for the structure in (d); (g) Same as in (a) for a sub-wavelength metal grating with sub-wavelength slits; (h) Reflection (blue, solid line), transmission (black, dotted line) and absorption (red, dashed line) vs. incident wavelength for the structure in (g); and (i) Reflected (blue, solid line), transmitted (black, dotted line), and absorbed (red, dashed line) SH conversion efficiency vs. pump/SH wavelengths for the structure in (g).

Finally, we consider a scenario in which a metallic grating with sub-wavelength slits (a=32nm and t=200nm) has sub-wavelength periodicity (p= 200 nm) [Fig. 3(g)]. Under these circumstances the only observable resonances in the linear spectra are those associated with the Fabry-Perót modes of the slits [r1, r2 and r3 in Fig. 3(h)]. These resonances are broad and are characterized by strong field localization in the slits [22] and lead to extraordinary optical transmission [23]. It is apparent that structures that support this kind of resonance tend to inhibit SH radiation and create broadband SH absorption: the nonlinear spectra always display an absorbed SH between two and three orders of magnitude larger than the emitted signal [Fig. 3(i)].

## 3. Discussion

In the previous sections we have not emphasized SH conversion efficiency *per se* since we focused on the ability of the structures to either emit or absorb SH light, and the ratio between these two quantities for different kinds of resonant features. In this section we attempt to summarize the results presented so far and provide a simple representation of the structure's performance in terms of two parameters: average electric field enhancement one the metal surface (FE) and a figure of merit that takes into account the ratio of the far-field radiated and the absorbed SH signals (FOM). These two quantities are thus defined as follows:

$$\mathrm{FE} = \mathrm{avg}\left(\left\|\mathbf{E}_{FF}\right\|^2 / \left\|\mathbf{E}_{FF,i}\right\|^2\right)_{metal}, \qquad (5)$$

$$\mathrm{FOM} = \eta_{SH}^2 / \alpha_{SH}, \tag{6}$$

where $\mathbf{E}_{FF}$ is the FF electric field on the metallic surface and $\mathbf{E}_{FF,i}$ is the incident electric field. In Eq. (5) the field enhancement is averaged on the metal surface. In Eq. (6), $\eta_{SH}$ is the emitted SH conversion efficiency (depending on the structure it is either the scattered or the sum of transmitted and reflected $P_{SH}$, normalized by the incident pump power $P_{FF}$) and $\alpha_{SH}$ is the absorbed SH signal, normalized by the incident pump power $P_{FF}$. For isolated nanostructures $P_{FF} = I_{FF} \cdot s \cdot \cos(\vartheta_i)$, where s is the geometrical cross section of the structure, whereas for periodic structures $P_{FF} = I_{FF} \cdot p \cdot \cos(\vartheta_i)$, where p is the periodicity. The ratio between $\eta_{SH}$ and $\alpha_{SH}$ is then multiplied by $\eta_{SH}$ so that the FOM takes into account the actual conversion efficiency of the structure. The FOM thus evaluates the ability of the structure to radiate, rather than absorb, at the SH frequency, simultaneously providing a measure of the structure's conversion efficiency in the far field, since this is the ultimate parameter that can be more easily measured in typical experimental settings.

A summary of the performance of all the structures we have evaluated is presented in Table 1. There we report the values for FE and FOM for two resonances (points A and B are indicated in all nonlinear spectra reported in the previous sections – Fig. 1(c), (f) and (i), Fig. 2(c), (f) and (i), Fig. 3(c), (f) and (i)). Table 1 also shows the order of magnitude of the radiated conversion efficiencies and a symbol associated with each structure, which will be used to provide a graphical representation of the structures' parameters. From a cursory examination of Table 1 one may infer that there are several structures that are quite similar in terms of emitted conversion efficiency (Kretschmann configuration, and point A for the isolated silver nanorod, coupled nanorods, nanorod arrays and periodic shallow grooves). It is also easy to verify that the emission efficiency achieved by pumping the structure at the lower resonance frequency is always larger compared to pumping at higher resonance frequencies. This behavior may be attributed to lower pump absorption losses due to poor penetration in the metal at lower frequencies.

Table 1. Summary of the structures' performance. Columns indicate – from left to right - : type of structure, FE for points A and B, FOM for points A and B, radiated conversion efficiency $\eta_{SH}$ for points A and B and symbols associated with each structure.

| Structure | FE | | FOM | | $\eta_{SH}$ | | Symbol |
|---|---|---|---|---|---|---|---|
| | A | B | A | B | A | B | |
| Mirror [Fig.1(a)] | 0.15 | - | $4.66 \cdot 10^{-9}$ | - | $10^{-9}$ | - | ▲ (red) |
| Thin Layer [Fig.1(d)] | 0.07 | - | $3.65 \cdot 10^{-9}$ | - | $10^{-10}$ | - | ▲ (blue) |
| Kretschmann [Fig.1(g)] | 2 | - | $7.1 \cdot 10^{-8}$ | - | $10^{-7}$ | - | △ |
| Nanorod [Fig.2(a)] | 1 | 5.5 | $1.52 \cdot 10^{-6}$ | $3.48 \cdot 10^{-11}$ | $10^{-7}$ | $10^{-10}$ | ● |
| Coupled Nanorods [Fig.2(d)] | 0.95 | 3.75 | $4.25 \cdot 10^{-8}$ | $1.25 \cdot 10^{-9}$ | $10^{-7}$ | $10^{-8}$ | ■ |
| Nanorods Array [Fig.2(g)] | 3.8 | 2.3 | $2.15 \cdot 10^{-7}$ | $0.8 \cdot 10^{-9}$ | $10^{-7}$ | $10^{-9}$ | ★ |
| Grooves [Fig.3(a)] | 1.25 | 0.06 | $1.4 \cdot 10^{-8}$ | $1.08 \cdot 10^{-9}$ | $10^{-7}$ | $10^{-9}$ | ◆ |
| Grating [Fig.3(d)] | 0.43 | 0.34 | $5.06 \cdot 10^{-9}$ | $2.4 \cdot 10^{-8}$ | $10^{-8}$ | $10^{-8}$ | ⬟ |
| Sub-wavelength Grating [Fig.3(g)] | 0.08 | 0.08 | $1 \cdot 10^{-10}$ | $1.9 \cdot 10^{-11}$ | $10^{-9}$ | $10^{-10}$ | ⬢ |

In order to evaluate the ability of each structure to either favor or impair SH emission we provide an accompanying graphical representation of two quantities listed in Table 1: FE and FOM. More specifically, in Fig. 4(a) we represent each structure described in the previous sections through a different symbol (see Table 1 for structure/symbol association). FE quantifies the ability of the structure to localize the field in the metal region, while FOM takes into account the actual capability of the structure to either radiate or absorb the SH signal, weighted by its actual radiated SH conversion efficiency. The graphical representation in Fig. 4(a) provides several clues that simply promoting local field enhancement and light confinement, and thus, strong surface-plasmon coupling, may not necessarily lead to efficient nonlinear frequency conversion. Indeed, among all structures having large FOMs, we find that the silver layer in the Kretschmann configuration (white triangle) has a FOM comparable to the coupled nanorods, nanorod arrays, or arrays with grooves and slits even with a relatively low FE. Another important aspect that influences the value of the FOM is the type of resonance under investigation. For example, an isolated nanorod can be either the most (green circle marked with A in Fig.4) or one of the least (green circle marked with B in Fig.4) convenient structures for nonlinear conversion efficiency, depending on the pumping frequency.

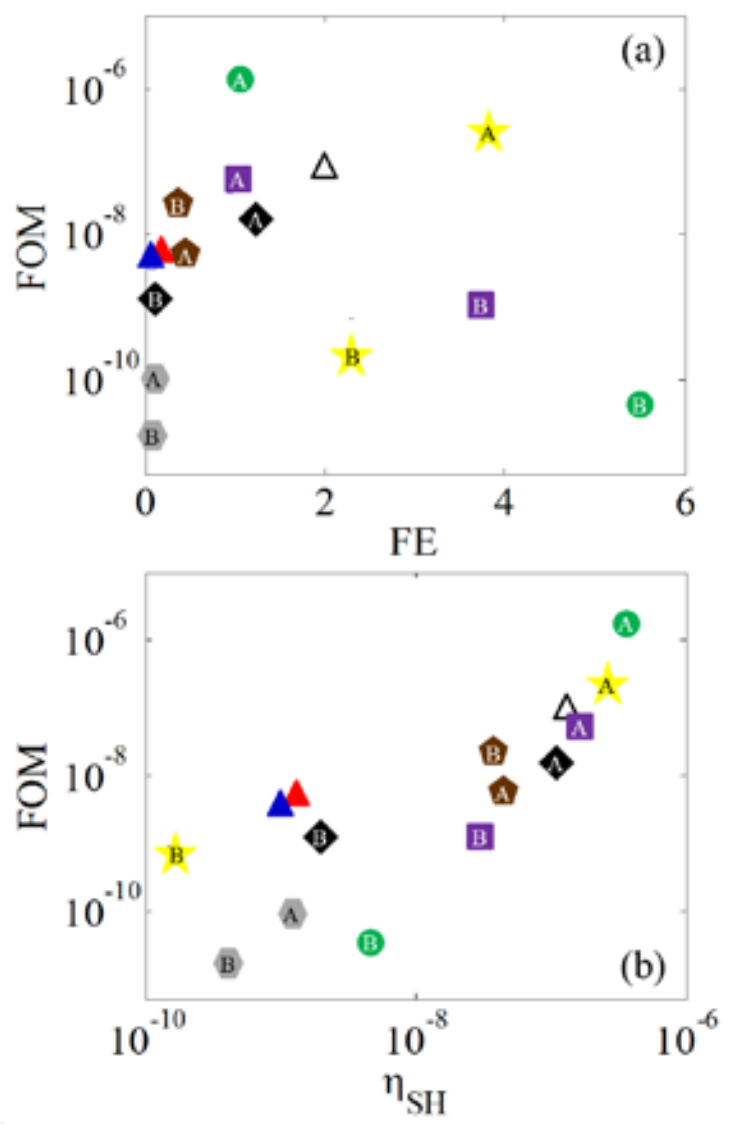

Figure 4. (a) FE vs FOM; (b) $\eta_{SH}$ vs FOM. Association of symbol and structures is presented in Table 1. Points A and B are indicated inside each symbol, where applicable.

More specifically, point A for the isolated nanorod corresponds to the excitation of a dipolar resonance for the pump and a quadrupolar resonance for the SH (see Fig. 2(b) and (c) - m1+m1→m2). This specific condition favors SH radiation, as the spectrum in Fig. 2(c) suggests. On the other hand, pumping the same structure at point B excites a quadrupolar resonance both at the pump and SH frequencies, resulting in one of the most unfavorable conditions for SH radiation and FOM (see Fig. 2(b) and (c) – m2+m2→m6). Similar conclusions may be reached for the coupled nanorods and array of nanorods (purple squares and yellow stars, respectively).

The situation is different for gratings with shallow grooves and slits. These structures are able to support Fano and Fabry-Perót-like resonances, or a combination of both [22, 61]. While the presence of an isolated Fano resonance only appears as a sharp modulation in the linear spectrum [see Fig. 3(b)], the interaction of these two resonances promotes the formation of a plasmonic band gap [22], shown in Fig. 3(e). However, all these structures are characterized by very relatively low values of FE, comparable with structures where plasmon excitation is actually absent, such as the silver mirror and thin layer. On the other hand, the FOM is high for gratings supporting Fano resonances and extremely low for Fabry-Perót-like resonances. These observations lead us infer that grating structures may either favor or impair SH radiation depending on the type of resonance at the FF.

The schematic representation in Fig. 4(a) generally reveals that by promoting only field enhancement does not necessarily lead to efficient SH emission. Indeed, there are several circumstances where an extremely low FE is paired with a high FOM, even when plasmonic resonances are not involved at all, like silver mirrors [red triangle in Fig. 4(a)] or thin silver films [blue triangle in Fig. 4(a)]. However, what appears to be more important from the analysis presented in the previous sections, and from its representation on the map in Fig. 4(a), is that the type of resonance present at either the pump and/or SH frequencies plays a major role in the radiation process: for example, dipolar resonances at the pump wavelength lead to higher SH conversion efficiencies when coupled with a quadrupolar resonance at the SH frequency. A similar observation can be done for metallic gratings, where Fano resonances favor SH radiation over absorption. These considerations may certainly be extended to metallic nanostructures with other shapes as well as metasurfaces and metamaterials.

As an alternative representation we also show in Fig. 4(b) how the FOM relates to the radiated conversion efficiency $\eta_{SH}$. Even though the definition of FOM establishes a quadratic relation with $\eta_{SH,}$ the representation in Fig. 4(b) allows immediately picturing the structures that would perform better in an experimental setting, placing the isolated nanorod (green circle – point A), the nanorods array (yellow star – point A), the silver layer in the Kretschmann configuration (white triangle) and the coupled nanorods (purple square – point A) among the best performing structures for second harmonic radiation.

While FOM in metallic nanostructures can be only improved by increasing the ability to radiate the SH signal, the same quantity can be boosted also by reducing $\alpha_{SH}$. A possible path to reduce the damping at the SH frequency is to use nanoresonators with low-loss nonlinear materials. Indeed all-dielectric nanostructures [62-64] and heterogeneous plasmonic composites [65, 66] are being investigated for harmonic generation to try to overcome the limits of all-metallic platforms. More specifically, all-dielectric nanostructures achieve efficient harmonic generation by exploiting a field localization that occurs mostly in the volume of the resonators. Under these circumstances the balance between the damping of the materials at the SH frequency and the increased field localization in the volume is crucial to determine the order of magnitude of $\alpha_{SH}$ and, therefore, the FOM.

## 4. Conclusions

A detailed analysis of nonlinear second order processes associated with a wide set of metallic nanostructures reveals that absorbed SH energy can be far greater than emitted SH energy, especially when coupling with surface plasmon is strong. By mapping the performances of all structures as a function of FE and FOM we found that indeed there is not a direct relation between the enhancement of the electric field inside the metallic region (FE) and the ability of the structure to radiate the SH signal in the far field (FOM). On the other hand, a more careful analysis that takes into account the type of resonance excited both at the pump and SH frequencies for each structure gives better insight on the structure's ability to radiate. However, we found that structures not intended to improve nonlinear frequency conversion might indeed be more favorable platforms for nonlinear optics. Among those, a silver mirror, a thin silver layer either in air or in the Kretschmann configuration actually show higher FOM with modest FEs. Our results therefore suggest caution should be exercised when designing plasmonic structures for frequency mixing purposes, paying attention to the types of resonances available and choosing pumping conditions accordingly.

**Acknowledgment**.
This research was performed while the authors M. A. Vincenti and D. de Ceglia held a National Research Council Research Associateship award at the U.S. Army Aviation and Missile Research Development and Engineering Center.